# Surface-induced positive planar Hall effect in topological Kondo insulator SmB$_6$ microribbons

L. Zhou[1], B. C. Ye[1], H. B. Gan[2], J. Y. Tang[1], P. B. Chen[1], Z. Z. Du[1], Y. Tian[2], S. Z. Deng[2], G. P. Guo[3], H. Z. Lu[1], F. Liu[2,*], and H. T. He[1,†]

[1]*Institute for Quantum Science and Engineering and Department of Physics, Southern University of Science and Technology, Shenzhen 518055, China*

[2]*State Key Laboratory of Optoelectronic Materials and Technologies, Guangdong Province Key Laboratory of Display Material and Technology, and School of Electronics and Information Technology, Sun Yat-sen University, Guangzhou 510275, China*

[3]*Key Laboratory of Quantum Information, CAS University of Science and Technology of China, Hefei 230026, China*

**Abstract**. Whether the surface states in SmB$_6$ are topological is still a critical issue in the field of topological Kondo insulators. In the magneto-transport study of single crystalline SmB$_6$ microribbons, we have revealed a positive planar Hall effect (PHE), the amplitude of which increases dramatically with decreasing temperatures but saturates below 5 K. This positive PHE is ascribed to the surface states of SmB$_6$ and expected to arise from the anisotropy in lifting the topological protection from back-scattering by the in-plane magnetic field, thus suggesting the topological nature of surface states in SmB$_6$. On the contrary, a negative PHE is observed for the bulk states at high temperatures, which is almost three orders of magnitudes weaker than the surface-induced positive PHE.

---

* liufei@mail.sysu.edu.cn
† heht@sustc.edu.cn

Topological insulators (TIs) hosting massless surface Dirac fermions have been extensively studied in the fields of condensed matter physics and material science for their profound exotic physics and potential applications in low-power spintronics or fault-tolerant quantum computing.[1,2] This novel TI phase is usually discovered in materials with strong spin-orbit couplings but weak electron correlations, such as $Bi_2Se_3$ and $Bi_2Te_3$.[3,4] But recent theoretical studies predicted a new class of TIs in certain Kondo insulators with strong electron correlations, termed as topological Kondo insulator (TKI).[5,6] A strong candidate of TKI is $SmB_6$, where topological surface states (TSSs) are expected to emerge in the bulk insulating gap due to the hybridization of even-parity *f*-electrons with odd-parity *d*-electrons. Compared with TIs, where the study of TSSs is often hindered by residual bulk states, a unique advantage of TKI is the dominant surface state conduction at low temperatures, manifested as the saturation of resistivity typically below 5 K.[6]

The existence of surface states in $SmB_6$ has been confirmed in many previous transport studies, including the capacitive self-oscillation experiment,[7] the point-contact spectroscopy,[8] the non-local and thickness independent transport measurement.[9,10] ARPES measurement also reveals the three surface Dirac cones located at the Γ and two X points, respectively.[11] Despite much progress, whether the surface states in $SmB_6$ are topological is still an open question in the field of TKI. The observed weak anti-localization effect indicates the possible presence of π Berry's phase in $SmB_6$.[12] The time reversal symmetry is also found critical in protecting the surface conduction in $SmB_6$, since the magnetic doping of $SmB_6$ can lead to the divergence of resistivity at low temperatures.[13] Furthermore, spin-resolved AREPS studies have revealed the linear energy dispersion and possible helical spin texture of surface states in $SmB_6$.[14,15] All these seem to indicate the topological nature of surface states in $SmB_6$. But a latest ARPES study claims that $SmB_6$ is a trivial surface conductor.[16] Therefore, more experimental evidence, especially the transport one, is highly desirable to verify the topological nature of surface states in $SmB_6$.

As enlightened by the latest studies of planar Hall effect (PHE) in TIs and topological semimetals,[17-22] which is found closely associated with the helical spin texture of TSSs in TIs or the chiral anomaly in semimetals, we have performed systematic study of PHE in TKI $SmB_6$.

At high temperatures with the bulk-state transport dominant, the PHE is negative and very weak. But when the bulk hybridization gap opens and the TSSs begin to emerge below about 60 K, the PHE changes its sign and shows a dramatic increase. The saturation of PHE is further observed below 5 K when the TSSs eventually dominate the transport in $SmB_6$. All these phenomena can be well understood in terms of the momentum-selective lifting of topological protections from back scattering in spin-helical TSSs, thus providing new transport evidence for the topological nature of TSSs in TKI $SmB_6$.

The $SmB_6$ samples we studied in this work are single-crystalline microribbons grown by chemical vapor deposition.[23-25] The growth axis of the ribbon is in the [110] direction, as indicated in the inset of Fig. 1 (a). To investigate the transport properties of $SmB_6$, six-terminal Hall-bar devices were fabricated by standard e-beam lithography and thermal evaporation techniques. The inset of Fig. 1 (a) shows the optical microscopy image of such a micro-device, with the Hall-bar width and length of 1.8 and 8.8 μm, respectively. We then measured these devices in a Quantum Design PPMS system using the lock-in technique. The amplitude and frequency of the ac current is 30 μA and 357 Hz, respectively. Since similar results were obtained in our $SmB_6$ microribbons, we only present the data of a 130-nm-thick sample.

Fig. 1 (a) shows the temperature ($T$) dependence of resistivity ($\rho$) of $SmB_6$. As temperatures decreases below about 60 K, a sharp increase of resistivity is clearly observed, reflecting the opening of the bulk energy gap due to the hybridization between the Sm $4f$ and $3d$ electrons.[26] More interestingly, the resistivity does not increase to infinity, but tends to saturate below about 5 K. Such a puzzling saturation of resistivity was previously ascribed to the impurity band conduction.[27] But since the discovery of TKI phase in $SmB_6$, it can be well interpreted by the dominant surface conduction emerging when the bulk hybridization gap fully opens.[5] Note that the magnetic doping can destroy this surface conduction, indicating the importance of the time reversal symmetry in TKI $SmB_6$.[13] Therefore, the saturation of resistivity below 5 K can be regarded as a transport signature of TSSs in TKI $SmB_6$.

We also measured the ordinary Hall effect (OHE) with the magnetic field ($B$) perpendicular to the surface of $SmB_6$ microribbons. Fig. 1 (b) shows the field-dependent Hall resistivity ($\rho_{OHE}$) at different temperatures. All the obtained $\rho_{OHE}(B)$ curves are linear, but there is a clear sign change in the Hall resistivity when the temperature decreases below 60 K. From the slope of the linear $\rho_{OHE}(B)$ curves, we can further deduce the sheet carrier density $n_H$. The temperature dependence of $n_H$ is plotted in Fig. 1 (a). In comparison with the $\rho(T)$ curve, one can see that the carrier changes from holes to electrons right around the temperature when the resistivity begins to increase rapidly. With the temperature further decreasing, there is a sharp drop of $n_H$ by almost 3 orders of magnitudes. Below about 10 K, the sheet carrier density $n_H$ saturates. Similar temperature dependence of $n_H$ has been observed in previous ARPES studies of SmB$_6$ crystals, showing that the Fermi level begins to locate in the bulk insulating gap around 60 K when the carrier changes its type.[28] Therefore, with temperatures decreasing below 60 K, the surface state gradually contributes to the transport in SmB$_6$ and finally shows its dominance as a saturation of resistivity below about 5 K.

We then investigated the PHE in SmB$_6$ by measuring the Hall resistivity in an in-plane magnetic field. The PHE arises from the anisotropy in the resistivity tensor and the planar Hall resistivity ($\rho_{PHE}$) can be phenomenologically described by the equation[29]

$$\rho_{PHE}(\theta, B) = \Delta\rho_{PHE}(B)sin(2\theta) = \frac{1}{2}(\rho_\parallel - \rho_\perp)sin(2\theta) \quad (1)$$

where $\rho_\parallel$ or $\rho_\perp$ is the longitudinal resistivity measured with the in-plane field parallel with or perpendicular to the measuring current $J$, $\Delta\rho_{PHE}$ is the amplitude of PHE, and $\theta$ is the angle of the field $B$ with respect to $J$, as schematically illustrated in the inset of Fig. 2 (b). From Eq. (1), the PHE is periodic in $\theta$ with a period of $\pi$ and is thus an even function of $B$, very different from the antisymmetric OHE. In our measurement of PHE, the field might not be perfectly in-plane. Therefore, the finite perpendicular component of the field would give rise to a small OHE signal. But due to the different symmetries between PHE and OHE, we can eliminate this signal by taking the average of the data in both positive and negative fields.[17]

Fig. 2 (a) shows the measured $\rho_{PHE}$ as a function of $\theta$ at $T$ =1.6 K and $B$ =14 T. The data can be well fitted by Eq. (1), as indicated by the solid fitting curve with the PHE amplitude $\Delta\rho_{PHE}$ =2.04E-7 Ωm. Furthermore, the obtained PHE amplitude exhibits a very strong temperature dependence, as shown in Fig. 2 (b). Below 5 K, the $\Delta\rho_{PHE}$ saturates. But as $T$ increases, it drops by almost three orders of magnitudes from 2.04E-7 Ωm at 1.6 K to 1.92E-10 Ωm at 50 K, resembling the temperature dependence of $\rho$. Besides this, the PHE amplitude even changes its sign around 60 K, *i.e.*, it is positive with $\rho_\parallel > \rho_\perp$ below 60 K (red color), but negative with $\rho_\parallel < \rho_\perp$ above 60 K (blue color). Fig. 2 (c) shows the angle dependence of $\rho_{PHE}$ at *T*=80 K and *B*=14 T. Compared with the $\rho_{PHE}(\theta)$ curve at 1.6 K in Fig. 2 (a), the PHE amplitude at 80 K (3.85E-10 $\Omega$m) is not only about three orders of magnitudes smaller, but also has an opposite sign. As discussed in Fig. 1, the bulk states dominate the transport of SmB$_6$ above 60 K. But with the temperature decreasing below 60 K, the surface states in SmB$_6$ become more important, and finally dominate the transport below 5 K. Therefore, the negative PHE observed above 60 K should be associated with the bulk states, while the positive PHE, which shows a rapid increase below 60 K and finally saturates below 5 K, is expected to arise from the surface states.

In addition to the sign, the PHE is also found to have different field dependence at high and low temperatures, respectively. Fig. 3 (a) shows the $\rho_{PHE}(\theta)$ curves measured at 1.6 K, but with different in-plane fields. It is clear that the PHE amplitude is greatly enhanced in higher fields. Fig. 3 (b) shows the field dependence of the obtained $\Delta\rho_{PHE}$ at 1.6 K. Note that the PHE amplitude has been normalized to the value of it at 14 T, *i.e.*, $\Delta\rho_{PHE}/\Delta\rho_{PHE}(14\,\text{T})$. Interestingly, the data can be well fitted by a power law $B^\alpha$ with $\alpha \sim 1.81$, as indicated by the red solid fitting curve in Fig. 3 (b). We have performed similar fitting process at other temperatures. The obtained value of $\alpha$ is plotted as a function of *T* in the inset of Fig. 3 (b). One can see that $\alpha$ is close to 1.8 for the surface-induced positive PHE observed with *T* below 60 K. But when the temperature is increased above 60 K, $\alpha$ is closer to 2 for the bulk negative PHE. Such a difference in the value of $\alpha$ also demonstrates the different physical origins of the positive and negative PHE we observed in SmB$_6$.

PHE is usually observed in ferromagnets due to the anisotropic *s-d* scattering and the sign of it could be positive or negative depending on the detailed *s-d* scattering process.[29,30] But the SmB$_6$ ribbons we studied are not magnetically doped. The linear OHE shown in Fig. 1 (b) also excludes the possible contribution of ferromagnetism to the observed PHE. Recently, topological semimetals have been shown to exhibit an unexpected giant PHE, which is closely associated with the intriguing phenomena of chiral anomaly.[18-22] Since the chiral charge pumping between different Weyl nodes will be suppressed when the magnetic field is tilted away from the current, $\rho_\perp > \rho_\parallel$ is expected. Therefore, the chiral anomaly induced PHE is negative, opposite to the positive PHE we observed in $SmB_6$ below 60 K when the surface states gradually dominate the transport.

Besides the ferromagnets and topological semimetals, we also notice the recent observation of PHE in TIs.[17] As is well known, the surface states in TIs are topologically protected from backscattering due to the spin-helical texture of TSSs.[1,2] But if an in-plane field is applied, this topological protection will be lifted anisotropically.[17] As schematically illustrated in Fig. 4 (a), when the in-plane field *B* is parallel with the applied current *J*, the spin-flip backscattering is allowed due to the time reversal symmetry breaking by the field. But if the field is perpendicular to the current in Fig. 4 (b), the backscattering is still topologically prohibited. As a result, one can expect $\rho_\parallel > \rho_\perp$ or positive PHE for TSSs in TIs. This positive PHE is intimately related to the helical spin texture of TSSs, thus reflecting the topological nature of TIs. As discussed in Fig. 1 & 2, the positive PHE which increases rapidly at low temperatures and tends to saturate below 5 K is most likely ascribed to the surface states in $SmB_6$. Furthermore, this surface-induced PHE has the same positive sign as that of TIs, *i.e.*, $\rho_\parallel > \rho_\perp$. These results lead us to believe that the positive PHE we observed also arises from the anisotropically lifting of the topological protection in the TSSs of $SmB_6$. Therefore, the observation of surface-induced positive PHE might provide a new transport evidence for the topological nature of surface states in TKI. As also revealed in Ref. 17, the in-plane magnetic field will spin polarize impurities in TIs, leading to more enhanced spin-flip impurity scattering in higher fields (see Fig. 4 (a)). This explains the strong field dependence of PHE shown in Fig. 3. But up to now, there is no theoretical model capable of explaining the observed power law with α~1.8 for the positive PHE, which certainly deserves further attention in the study of TKI or TI.

Note that the positive PHE has also been reported in a previous study of $SmB_6$ polycrystalline films with nanometer-sized grains.[31] But the magnitude of the positive PHE is almost two orders of magnitudes smaller than that we obtained in single crystalline $SmB_6$ microribbons. In that work, the positive PHE can only be observed in polycrystalline films with thickness smaller than 32 nm and the physical origin of it was ascribed to the hybridization between the top and bottom surface states of $SmB_6$.[31] These are very different from our results, since the positive

PHE can be observed in single crystalline SmB$_6$ ribbons with thickness up to 130 nm, where the coupling between the top and bottom surface states certainly cannot occur. Therefore, the most possible mechanism for the positive PHE in single crystalline SmB$_6$ is the anisotropy in lifting the topological protection from back scattering, as we just discussed above. At present, we have no idea about the mechanism for the negative PHE observed above 60 K, which is almost three orders of magnitudes weaker than the positive PHE and believed to arise from the bulk states in SmB$_6$. More efforts are needed in the near future to clarify this issue.

In conclusion, we have investigated the PHE in single crystalline SmB$_6$ microribbons. A negative PHE is observed above 60 K for the bulk states of SmB$_6$. But below 60 K, a positive PHE emerges, the magnitude of which shows a rapid increase with decreasing temperatures and finally saturates below 5 K. This positive PHE is ascribed to the surface states of SmB$_6$ and can be well interpreted in terms of the anisotropy in lifting the topological protection by the in-plane magnetic field. Therefore, our study reveals a new TSS-related transport phenomenon in TKI SmB$_6$ and provides further transport evidence for the helical spin texture in the TSSs of SmB$_6$.

**ACKNOWLEDGEMENTS**

This work was supported by the National Natural Science Foundation of China (No. 11574129, 51872337), the National Key Research and Development Program of China (No. 2016YFA0301703), National Project for the Development of Key Scientific Apparatus of China (No. 2013YQ12034506), the Fundamental Research Funds for the Central Universities of China, the Natural Science Foundation of Guangdong Province (No. 2015A030313840), and Technology and Innovation Commission of Shenzhen Municipality (No. KQJSCX20170727090712763).

**FIGURE CAPTIONS**

FIG 1. (a) Temperature dependent resistivity and sheet carrier density of $SmB_6$ microribbons. (b) Field dependent ordinary Hall resistivity at different temperatures. For clarity, the Hall resistivity with $T$ above 80 K has been multiplied by 100.

FIG 2. (a) Angle dependence of planar Hall resistivity measured with $B$=14 T and $T$ = 1.6 K. (b) Temperature dependence of the PHE amplitude and resistivity. Inset: definition of the tilting angle $\theta$. (c) Angle dependence of planar Hall resistivity measured with $B$=14 T and $T$ = 80 K.

FIG 3. (a) Angle dependence of the PHE resistivity at 1.6 K but different in-plane fields. (b) Magnetic field dependence of the normalized PHE amplitude at 1.6 K. The data is fitted by the power law $B^\alpha$, as indicated by the solid fitting curve. Inset: Temperature dependence of $\alpha$.

FIG 4. Topological protection from back scattering is lifted with $\mathbf{B} \parallel \mathbf{J}$ (a), but still preserved with $\mathbf{B} \perp \mathbf{J}$ (b).

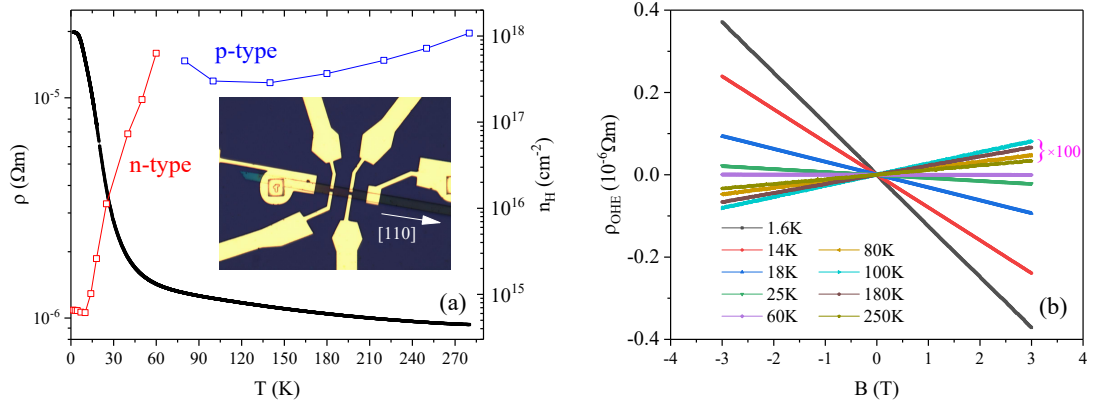

**Figure 1**

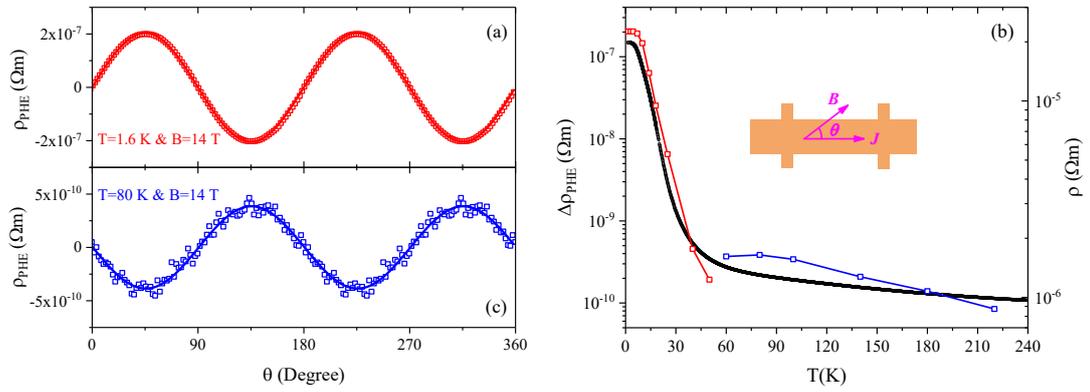

**Figure 2**

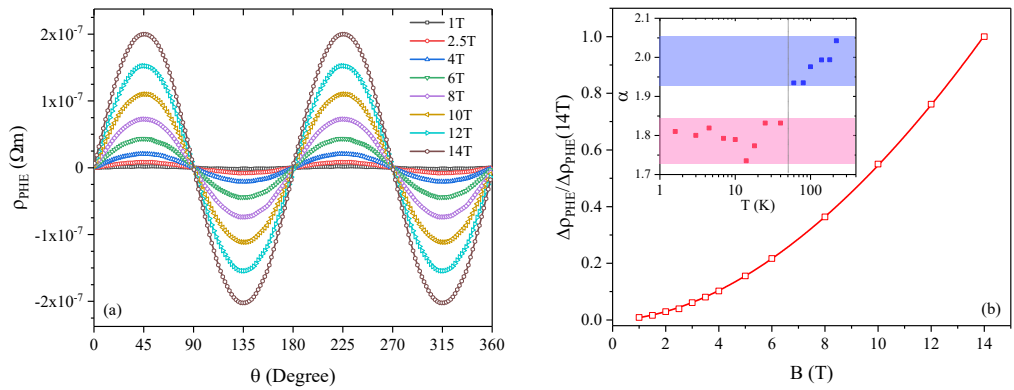

**Figure 3**

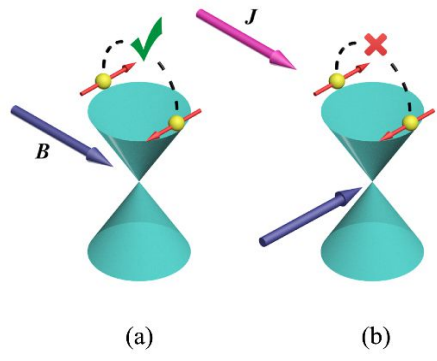

**Figure 4**